\documentclass{ieeeoj_mod}
\usepackage{cite}
\usepackage{amsmath,amssymb,amsfonts}
\usepackage{algorithmic}
\usepackage{graphicx,color}
\usepackage{textcomp}
\usepackage{hyperref}

\usepackage{algorithm}      
\usepackage{algorithmic}      
\usepackage{eqparbox}
\newdimen{\algindent}
\definecolor{commentcolor}{cmyk}{.46,.0,.50,.16}			          
\newcommand\doublerulefill{\leavevmode\leaders\vbox{\hrule width .1pt\kern1pt\hrule}\hfill\kern0pt }  
\def\BibTeX{{\rm B\kern-.05em{\sc i\kern-.025em b}\kern-.08em
    T\kern-.1667em\lower.7ex\hbox{E}\kern-.125emX}}
\AtBeginDocument{\definecolor{ojcolor}{cmyk}{0.93,0.59,0.15,0.02}}
\def\OJlogo{\vspace{-5pt}\includegraphics[height=16.5pt]{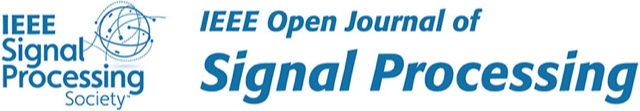}}

\DeclareMathOperator*{\argmax}{argmax}
\DeclareMathOperator*{\argmin}{argmin}
\makeatletter
\renewcommand*\env@matrix[1][*\c@MaxMatrixCols c]{%
\hskip -\arraycolsep
\let\@ifnextchar\new@ifnextchar
\array{#1}}
\makeatother

\begin{document}
\receiveddate{XX Month, XXXX}
\reviseddate{XX Month, XXXX}
\accepteddate{XX Month, XXXX}
\publisheddate{XX Month, XXXX}
\currentdate{XX Month, XXXX}
\doiinfo{OJSP.2023.1234567}

\title{Sampling Rate Offset Estimation and Compensation for Distributed Adaptive Node-Specific Signal Estimation in Wireless Acoustic Sensor Networks}

\author{PAUL DIDIER\IEEEauthorrefmark{1}, TOON VAN WATERSCHOOT\IEEEauthorrefmark{1}, SIMON DOCLO\IEEEauthorrefmark{2}, AND MARC MOONEN\IEEEauthorrefmark{1}}
\affil{STADIUS Center for Dynamical Systems, Department of Electrical Engineering (ESAT),\\KU Leuven, 3001 Leuven, Belgium}
\affil{Signal Processing Group, Department of Medical Physics and Acoustics and Cluster of Excellence Hearing4all,\\University of Oldenburg, Oldenburg, Germany}
\corresp{CORRESPONDING AUTHOR: Paul Didier (e-mail: paul.didier@esat.kuleuven.be).}
\authornote{This research work was carried out in the frame of the European Union's Horizon 2020 research and innovation programme under the Marie Skłodowska-Curie Grant Agreement No. 956369: ``Service-Oriented Ubiquitous Network-Driven Sound — SOUNDS''. The research leading to these results has received funding from the European Research Council under the European Union's Horizon 2020 research and innovation program / ERC Consolidator Grant: SONORA (no. 773268). This paper reflects only the authors' views and the Union is not liable for any use that may be made of the contained information. The scientific responsibility is assumed by the authors.}
\markboth{SRO ESTIMATION AND COMPENSATION FOR DISTRIBUTED ADAPTIVE NODE-SPECIFIC SIGNAL ESTIMATION IN WASNS}{Paul Didier \textit{et al.}}

\begin{abstract}
    Sampling rate offsets (SROs) between devices in a heterogeneous wireless acoustic sensor network (WASN) can hinder the ability of distributed adaptive algorithms to perform as intended when they rely on coherent signal processing. 
    In this paper, we present an SRO estimation and compensation method to allow the deployment of the distributed adaptive node-specific signal estimation (DANSE) algorithm in WASNs composed of asynchronous devices.
    The signals available at each node are first utilised in a coherence-drift-based method to blindly estimate SROs which are then compensated for via phase shifts in the frequency domain. A modification of the weighted overlap-add (WOLA) implementation of DANSE is introduced to account for SRO-induced full-sample drifts, permitting per-sample signal transmission via an approximation of the WOLA process as a time-domain convolution. 
    The performance of the proposed algorithm is evaluated in the context of distributed noise reduction for the estimation of a target speech signal in an asynchronous WASN.
\end{abstract}

\begin{IEEEkeywords}
  Sampling rate offsets, coherence drift, signal enhancement, weighted overlap-add, wireless acoustic sensor networks
\end{IEEEkeywords}


\maketitle

\section{Introduction}\label{sec:intro}

Wireless acoustic sensor networks (WASNs) have been a subject of great interest in recent years as they provide a number of advantages over centralised systems performing audio signal processing tasks~\cite{bertrand_applications_2011}.
Novel algorithmic solutions aim to utilise the increased flexibility and scalability of WASNs in order to tackle various audio signal processing challenges in a distributed fashion, bypassing the need for a data fusion centre with which all nodes communicate.

This paper focuses on distributed signal estimation, where each node in the WASN aims to estimate a node-specific desired signal.
The distributed adaptive node-specific signal estimation (DANSE) algorithm was originally formulated in~\cite{bertrand_distributed_2010,bertrand_distributed_2010-1} to tackle this problem. This algorithm is designed to allow each node of a fully connected WASN to achieve centralised performance upon convergence by iteratively computing its own multichannel Wiener filter (MWF) while only exchanging single-channel signals with other nodes. DANSE can significantly reduce the required number of signals communicated between nodes in a WASN with many sensors per node, compared to a centralised MWF where nodes communicate with a single fusion centre.
Although the DANSE algorithm has been tested under various conditions and for different tasks~\cite{ruiz_distributed_2022}, a key aspect allowing its robust deployment in real-world scenarios has yet to be addressed, namely, signals asynchronicity. 

In many practical applications such as teleconferencing systems or smart domotics, the WASN is heterogeneous, i.e., composed of various interconnected devices such as laptops, tablets, or hearing aids.
Each device samples the incoming acoustic information at a specific rate via its own analog-to-digital converter based on an internal clock, the sampling rate of which may differ from the nominal value provided by the manufacturer~\cite{he_analysis_2015}.
The sampling rate mismatch between two devices can be quantified as the sampling rate offset (SRO), generally expressed in parts-per-million (PPM).
SROs in the range of $\pm$500 PPM have been measured between commonly used devices and reported in~\cite{he_analysis_2015}. The same study showed that SROs can slowly vary through time, e.g., when the devices undergo significant temperature changes or fluctuations in supply voltage.

SROs lead to an increasing time-drift between signals sampled by different clocks, which inhibits their use in algorithms that rely on coherent signal processing~\cite{lienhart_importance_2003}.
Notably, the performance of signal enhancement algorithms based on the MWF such as the DANSE algorithm depends on the computation of accurate spatial covariance matrices. DANSE can thus be expected to be sensitive to a lack of synchronicity between locally recorded microphone signals and signals received from other nodes. In fact, literature around DANSE has so far assumed that all nodes involved in the algorithm have exactly the same sampling rate~\cite{bertrand_distributed_2010,bertrand_distributed_2010-1,hassani_gevd-based_2016,szurley_topology-independent_2017,van_rompaey_distributed_2021}. The asynchronicity problem in WASNs has recently been investigated in the context of algorithms other than DANSE~\cite{zhang_joint_2020,de_hu_distributed_2023}.

In this paper, we propose a methodology to relax the synchronicity assumption in DANSE, bringing this algorithm closer to robust deployment in real-life scenarios.
The presence of SROs is addressed in a fully connected WASN where node and source positions are fixed. Time-invariant SROs are considered, assuming that no temperature or supply voltage changes occur during the convergence phase of the algorithm.
Per-node-pair SRO estimation is performed blindly based on a coherence-drift method~\cite{schmalenstroeer_multi-stage_2017, gburrek_synchronization_2022}.
The weighted overlap-add (WOLA) implementation of the generalised eigenvalue decomposition (GEVD-)DANSE algorithm~\cite{bertrand_robust_2009} is modified to permit detection of full-sample drifts (FSDs) via per-sample signal broadcasting. This is achieved by approximating the WOLA process used for local signal fusion (analysis, filtering in the short-time Fourier transform (STFT) domain, and synthesis) as a single time-domain convolution operation. This method allows to retain the low complexity of WOLA processing for the more costly steps of GEVD-based filter update and desired signal estimation.
The estimated SROs and the detected FSDs are then compensated for via phase shifts in the STFT-domain. 
The performance of the proposed algorithm is evaluated in the context of distributed noise reduction for the estimation of a target speech signal.

The paper is organised as follows. In Section~\ref{sec:MWF}, the centralised GEVD-MWF is reviewed. The key aspects of the theory and implementation of the DANSE algorithm are summarised in Section~\ref{sec:danse}. The proposed method for SRO estimation and compensation within the DANSE framework is presented in detail in Section~\ref{sec:sroestcomp}. The performance of the proposed method is then analysed by means of simulations in asynchronous WASNs in Section~\ref{sec:results}. Finally, conclusive remarks are formulated in Section~\ref{sec:ccl}.

\section{GEVD-MWF-based signal estimation}\label{sec:MWF}

A WASN composed of $K$ nodes is considered, where each node $k\in\mathcal{K}=\{1,\hdots,K\}$ has $M_k \geq 1$ microphones. The total number of microphones in the network is denoted by $M=\sum_{k\in\mathcal{K}} M_k$.
In the acoustic scene, one localised static desired signal source (e.g., a talker) and $J\geq 1$ localised static noise sources are present. The signals recorded by node $k$ can be represented in the STFT domain at frame $i$ and frequency bin $\nu$ via an additive-noise signal model:

\begin{equation}\label{eq:sigModel_TD}
  \mathbf{y}_k[\nu,i] = \mathbf{s}_k[\nu,i] + \mathbf{n}_k[\nu,i],
\end{equation}

\noindent
where $\mathbf{y}_k[\nu,i]$, $\mathbf{s}_k[\nu,i]$, and $\mathbf{n}_k[\nu,i]$ are $M_k$-dimensional vectors corresponding to the microphone signals, the desired signal components of these signals, and the noise components, respectively. The additive noise is assumed to be uncorrelated with the desired signal.

In centralised processing, the signal vector available at the fusion centre is defined as an $M$-dimensional stacked version  $\mathbf{y}[\nu,i] = \begin{bmatrix}
  \mathbf{y}_1^T[\nu,i],\dots,\mathbf{y}_K^T[\nu,i]
\end{bmatrix}^T$ of the node-specific microphone signals where $(\cdot)^T$ denotes the transpose operation. Similarly to~\eqref{eq:sigModel_TD}, this vector can be expressed as $\mathbf{y}[\nu,i] = \mathbf{s}[\nu,i] + \mathbf{n}[\nu,i]$
with $\mathbf{n}[\nu,i] = \begin{bmatrix}
  \mathbf{n}_1^T[\nu,i],\dots,\mathbf{n}_K^T[\nu,i]
\end{bmatrix}^T$ and $\mathbf{s} = \begin{bmatrix}
  \mathbf{s}_1^T[\nu,i],\dots,\mathbf{s}_K^T[\nu,i]
\end{bmatrix}^T$. 

The objective of node $k$ is then to estimate a local desired signal $d_k[\nu,i]$ based on $\mathbf{y}[\nu,i]$. Define without loss of generality (w.l.o.g.) the desired signal $d_k[\nu,i]$ at node $k$ to be the desired signal component of the first local microphone signal, i.e., $d_k[\nu,i] = \mathbf{e}_{d_k}^T\mathbf{s}[\nu,i]$, where $\mathbf{e}_{d_k}$ selects the appropriate channel of $\mathbf{s}[\nu,i]$. 
An optimal filter $\bar{\mathbf{w}}_k[\nu,i]$ can be obtained by minimising the mean squared error (MSE) between the desired signal and the filtered microphone signals:

\begin{equation}\label{eq:MMSE}
  \bar{\mathbf{w}}_k[\nu,i+1] = \underset{\mathbf{w}_k[\nu]}{\argmin} \: E\left\{ \left| d_k[\nu,i] - \mathbf{w}_k^H[\nu]\mathbf{y}[\nu,i] \right|^2 \right\},
\end{equation}

\noindent
where $\left(\cdot\right)^H$ denotes complex conjugation and $E\{\cdot\}$ the expected value operation.
The closed-form solution of~\eqref{eq:MMSE} is the MWF:

\begin{equation}\label{eq:MWFbasic}
  \bar{\mathbf{w}}_k[\nu,i+1] = \left(\bar{\mathbf{R}}_{\mathbf{y}\mathbf{y}}[\nu,i]\right)^{-1} \bar{\mathbf{R}}_{\mathbf{s}\mathbf{s}}[\nu,i] \mathbf{e}_{d_k},
\end{equation}

\noindent
where $\bar{\mathbf{R}}_{\mathbf{y}\mathbf{y}}[\nu,i] = E\{ \mathbf{y}[\nu,i]\mathbf{y}^H[\nu,i] \}$ is the network-wide microphone signal covariance matrix and $\bar{\mathbf{R}}_{\mathbf{s}\mathbf{s}}[\nu,i] = \bar{\mathbf{R}}_{\mathbf{y}\mathbf{y}}[\nu,i] - \bar{\mathbf{R}}_{\mathbf{n}\mathbf{n}}[\nu,i]$, where $\bar{\mathbf{R}}_{\mathbf{n}\mathbf{n}}[\nu,i] = E\{ \mathbf{n}[\nu,i]\mathbf{n}^H[\nu,i] \}$ is the network-wide noise-only covariance matrix. Assuming short-term stationarity of the signals, the covariance matrices can be estimated by averaging over observations of $\mathbf{y}[\nu,i]\mathbf{y}^H[\nu,i]$. 
In a speech enhancement scenario with stationary noise, the on-off behaviour of the desired signal can be exploited via a voice activity detector (VAD)~\cite{bertrand_energy-based_2010,zhao_model-based_2020} to isolate noise-only observations of $\mathbf{y}[\nu,i]$. The estimation of $\bar{\mathbf{R}}_{\mathbf{y}\mathbf{y}}[\nu,i]$ and $\bar{\mathbf{R}}_{\mathbf{n}\mathbf{n}}[\nu,i]$ can then be performed via exponential averaging:

\begin{equation}\label{eq:VADexpavg}
  \begin{split}
    \text{VAD = 1 :}\\
    \hat{\mathbf{R}}_{\mathbf{y}\mathbf{y}}[\nu,i] &= \beta \hat{\mathbf{R}}_{\mathbf{y}\mathbf{y}}[\nu,i-1] + (1-\beta) \mathbf{y}[\nu,i]\mathbf{y}^H[\nu,i],\\
    \text{VAD = 0 :}\\
    \hat{\mathbf{R}}_{\mathbf{n}\mathbf{n}}[\nu,i] &= \beta \hat{\mathbf{R}}_{\mathbf{n}\mathbf{n}}[\nu,i-1] + (1-\beta) \mathbf{y}[\nu,i]\mathbf{y}^H[\nu,i],
  \end{split}
\end{equation}

\noindent
where the real-valued number $\beta$ acts as a forgetting factor, $0\ll \beta \leq 1$, typically chosen close to 1 to preserve spatial coherence between microphone signals~\cite{bertrand_applications_2011}.

In the presence of a single desired signal source, the signal model implies that $\bar{\mathbf{R}}_{\mathbf{s}\mathbf{s}}[\nu,i]$ should be a rank-1 matrix~\cite{serizel_low-rank_2014}. However, the estimated $\hat{\mathbf{R}}_{\mathbf{s}\mathbf{s}}[\nu,i] = \hat{\mathbf{R}}_{\mathbf{y}\mathbf{y}}[\nu,i] - \hat{\mathbf{R}}_{\mathbf{n}\mathbf{n}}[\nu,i]$ generally has a rank larger than 1. A GEVD-based approach was proposed in~\cite{serizel_low-rank_2014} to obtain a rank-1 approximation of $\hat{\mathbf{R}}_{\mathbf{s}\mathbf{s}}[\nu,i]$.
The GEVD of the matrix pencil $\{\hat{\mathbf{R}}_{\mathbf{y}\mathbf{y}}[\nu,i], \hat{\mathbf{R}}_{\mathbf{n}\mathbf{n}}[\nu,i]\}$ yields:

\begin{equation}\label{eq:gevd}
  \begin{split}
    \hat{\mathbf{R}}_{\mathbf{y}\mathbf{y}}[\nu,i] &= \hat{\mathbf{Q}}[\nu,i]\hat{\mathbf{\Sigma}}[\nu,i]\hat{\mathbf{Q}}^H[\nu,i],\\
    \hat{\mathbf{R}}_{\mathbf{n}\mathbf{n}}[\nu,i] &= \hat{\mathbf{Q}}[\nu,i]\hat{\mathbf{Q}}^H[\nu,i],
  \end{split}
\end{equation}

\noindent
with $\hat{\mathbf{Q}}[\nu,i]$ an $M\times M$ matrix of which the columns are the generalised eigenvectors (GEVCs) and $\hat{\mathbf{\Sigma}}[\nu,i]$ is a diagonal matrix of which the diagonal elements are the corresponding generalised eigenvalues (GEVLs). The GEVLs in $\hat{\mathbf{\Sigma}}[\nu,i]$ are assumed to be ordered by decreasing magnitude.
A rank-1 approximation of $\bar{\mathbf{R}}_{\mathbf{s}\mathbf{s}}[\nu,i]$ can then be obtained by using~\eqref{eq:gevd} and nullifying the $M-1$ smallest GEVLs.
Substituting into~\eqref{eq:MWFbasic} then leads to the GEVD-MWF:

\begin{equation}\label{eq:gevd-mwf}
  \hat{\mathbf{w}}_k[\nu,i+1] = \hat{\mathbf{Q}}^{-H}[\nu,i]\mathbf{\Lambda}[\nu,i]\hat{\mathbf{Q}}^H[\nu,i]\mathbf{e}_{d_k},
\end{equation}

\noindent
with $\mathbf{\Lambda}[\nu,i] = \mathrm{diag}\{ 1 - 1/\hat{\sigma}_1[\nu,i],0,\dots,0 \}$, where $\mathrm{diag}\{\cdot\}$ transforms a vector into a diagonal matrix and $\hat{\sigma}_1[\nu,i]$ is the largest GEVL. Finally, the desired signal at frequency bin $\nu$ and frame $i$ is estimated as $\hat{d}_k[\nu,i] = \hat{\mathbf{w}}_k^H[\nu,i+1] \mathbf{y}[\nu,i]$.

\section{The DANSE algorithm}\label{sec:danse}

The DANSE algorithm~\cite{bertrand_distributed_2010} provides a distributed implementation of the MWF described in Section~\ref{sec:MWF} as an adaptive algorithm where nodes iteratively update their local filter estimates.
Different node-updating schemes exist: (i) sequential updating~\cite{bertrand_distributed_2010}, (ii) simultaneous updating~\cite{bertrand_distributed_2010-1}, and (iii) asynchronous updating~\cite{bertrand_distributed_2010-1}. Strategies (i) and (ii) rely on a network-wide update protocol that coordinates the updates, unlike strategy (iii)~\cite{bertrand_distributed_2010-1}.
Since the presence of unknown SROs between nodes challenges the deployment of a coordination protocol, asynchronous updating is assumed in the following.
To avoid limit cycles due to asynchronous updating~\cite{bertrand_distributed_2010-1}, relaxed filter updates can be performed~\cite{szurley_improved_2013}.
Computational delays due to data transmission, reception, and processing are assumed to be negligible in this paper. 

As described in~\cite{bertrand_robust_2009}, the DANSE algorithm can be implemented using weighted overlap-add (WOLA) processing to efficiently perform short-time Fourier analysis and synthesis~\cite{crochiere_weighted_1980}.
Using WOLA, time-domain microphone signals are processed on a frame-by-frame basis. WOLA analysis consists of applying an $N$-point DFT to a windowed frame of a time-domain signal, with the frame size equal to the DFT size, effectively transforming the time-domain signal frame into an STFT-domain signal frame. In DANSE, all filtering can be conducted in the STFT domain, resulting in a lower computational complexity as compared to a time-domain implementation~\cite{bertrand_robust_2009}.
Each new WOLA frame then corresponds to a new DANSE iteration where the nodes update their filter estimate. As a WOLA implementation of DANSE is assumed in the following, the variable $i$ simultaneously denotes the STFT frame index as well as the DANSE iteration index, i.e., nodes update their filter estimates at each new $i$.

Although a variety of network topologies can exist, a fully connected WASN is assumed in this paper. All the nodes that can communicate with node $k$ are grouped in the set $\mathcal{K}_k = \mathcal{K}\backslash\{k\}$. 
The DANSE algorithm in a fully connected WASN operates in two main stages: signals fusion and broadcasting on the one hand, and filters updates on the other hand~\cite{bertrand_distributed_2010}. At each frame $i$, each node $k$ fuses its $M_k$ local microphone signals into a single-channel signal $z_k[\nu,i]$ $\forall\:\nu\in\{1\dots N\}$ using a fusion vector $\mathbf{p}_k[\nu,i]$ before broadcasting it to the other nodes, which reduces the per-node communication cost by a factor $M_k$ (i.e., a factor $M / K$ over the entire network) compared to the centralised MWF of Section~\ref{sec:MWF}. An appropriate definition of $\mathbf{p}_k[\nu,i]$ guarantees convergence of the DANSE algorithm to the centralised MWF solution~\cite{bertrand_distributed_2010}. 
In the WOLA implementation, a time-domain fused signal denoted by $\dot{z}_k[n]$ is obtained via WOLA synthesis (inverse DFT followed by windowing) and overlap-add of the fused signal frames $z_k[\nu,i]$, where $n$ denotes the sample index. The time-domain signal $\dot{z}_k[n]$ is then broadcast to other nodes, as summarised in Algorithm~\ref{alg:woladanse}.

The STFT-domain signals available at node $k$ at iteration $i$ are grouped into the vector:

\begin{equation}
  \tilde{\mathbf{y}}_k[\nu,i] = \begin{bmatrix}
    \mathbf{y}_k^T[\nu,i] \:|\: \mathbf{z}_{-k}^T[\nu,i]
  \end{bmatrix}^T
\end{equation}

\noindent
where $\mathbf{y}_k[\nu,i]$ contains the local microphone signals and $\mathbf{z}_{-k}[\nu,i]$ is a stacked version of all the microphone signals received from other nodes.
As in the centralised case (cfr.~\eqref{eq:sigModel_TD}), $\tilde{\mathbf{y}}_k[\nu,i]$ can be written as a sum of a desired signal component $\tilde{\mathbf{s}}_k[\nu,i]$ and a noise component $\tilde{\mathbf{n}}_k[\nu,i]$.
Node $k$ aims to compute the $i$-th STFT frame of its desired signal estimate $\hat{d}_k[\nu,i]$ via multichannel filtering of $\tilde{\mathbf{y}}_k[\nu,i]$. The filter at node $k$ is denoted by $\tilde{\mathbf{w}}_k[\nu,i] = [\mathbf{w}_{kk}^T[\nu,i] \:|\: \mathbf{g}_{k-k}^T[\nu,i]]^T$, where $\mathbf{w}_{kk}[\nu,i]$ is applied to the local microphone signals $\mathbf{y}_k[\nu,i]$ and $\mathbf{g}_{k-k}[\nu,i]$ is applied to the fused microphone signals $\mathbf{z}_{-k}[\nu,i]$.
The filter at node $k$ at iteration $i+1$ is obtained by minimising the MSE between the desired signal and its estimate:

\begin{equation}\label{eq:mmse_danse}
  \tilde{\mathbf{w}}_k[\nu,i+1] = \underset{\mathbf{w}_k[\nu]}{\argmin} \: E\left\{ \left| d_k[\nu,i] - \mathbf{w}_k^H[\nu]\tilde{\mathbf{y}}_k[\nu,i] \right|^2 \right\},
\end{equation}

\noindent
and the $i$-th STFT frame of desired signal estimate is then $\hat{d}_k[\nu,i] = \tilde{\mathbf{w}}_k^H[\nu,i+1] \tilde{\mathbf{y}}_k[\nu,i]$.
Equation~\eqref{eq:mmse_danse} has the same structure as~\eqref{eq:MMSE}, be it with a different definition of the filter $\mathbf{w}_k[\nu]$ and input vector $\tilde{\mathbf{y}}_k[\nu,i]$, hence its solution again corresponds to an MWF. With the covariance matrices $\tilde{\mathbf{R}}_{\mathbf{y}_k\mathbf{y}_k}[\nu,i]$ and $\tilde{\mathbf{R}}_{\mathbf{n}_k\mathbf{n}_k}[\nu,i]$ defined and estimated per node instead of centrally as in~\eqref{eq:VADexpavg}, a GEVD is applied to the matrix pencil $\{ \tilde{\mathbf{R}}_{\mathbf{y}_k\mathbf{y}_k}[\nu,i], \tilde{\mathbf{R}}_{\mathbf{n}_k\mathbf{n}_k}[\nu,i] \}$ and, similarly to~\eqref{eq:gevd-mwf}, the filter is computed as:

\begin{equation}\label{eq:danse_filter_update}
  \tilde{\mathbf{w}}_k[\nu,i+1] = \tilde{\mathbf{Q}}_k^{-H}[\nu,i]\tilde{\mathbf{\Lambda}}_k[\nu,i]\tilde{\mathbf{Q}}_k^H[\nu,i]\mathbf{e}_{d_k},
\end{equation}

\noindent
where, at frequency $\nu$ and iteration $i$, $\tilde{\mathbf{Q}}_k[\nu,i]$ is an $(M_k+K-1)\times (M_k+K-1)$ matrix of which the columns are the GEVCs, $\tilde{\mathbf{\Lambda}}_k[\nu,i] = \mathrm{diag}\left\{ 1 - 1/\tilde{\sigma}_{k1}[\nu,i], 0,\dots,0 \right\}$ and $\tilde{\sigma}_{k1}[\nu,i]$ is the largest GEVL.
Finally, to ensure convergence of $\mathbf{w}_{kk}[\nu,i]$ towards the corresponding elements of the centralised MWF, the fusion vector at iteration $i$ is defined as $\mathbf{p}_k[\nu,i] = \mathbf{w}_{kk}[\nu,i]$, such that:

\begin{equation}\label{eq:compressed_signal_danse}
  z_k[\nu,i] = \mathbf{w}_{kk}^H[\nu,i]\mathbf{y}_k[\nu,i].
\end{equation}

The WOLA implementation of the DANSE algorithm is summarised in Algorithm~\ref{alg:woladanse}, where the $M_k$ local time-domain microphone signals at node $k$ are denoted by $\dot{\mathbf{y}}_k[n]$. The time-domain signal obtained after WOLA synthesis and overlap-add of consecutive frames $\hat{d}_k[\nu,i]$ is denoted by $\dot{\hat{d}}_k[n]$. The time-domain fused signals are grouped in the vector $\dot{\mathbf{z}}_{-k}[n] = \begin{bmatrix}
  \dot{z}_1[n] \: \dots \: \dot{z}_{k-1}[n] \: \dot{z}_{k+1}[n] \: \dots \: \dot{z}_K[n]
\end{bmatrix}^T$. The WOLA window shift, corresponding to the number of new samples recorded between two consecutive DANSE iterations, is denoted by $N_\mathrm{s}$. 

\begin{algorithm}
  \caption{WOLA-based DANSE in a synchronised, fully connected WASN (50\% window overlap).}\label{alg:woladanse}
  \begin{algorithmic}[1]
    \STATE Initialise $\tilde{\mathbf{w}}_k[\nu,0]$ $\forall\:k\in\mathcal{K}$;
    \STATE Each node $k\in\mathcal{K}$ performs, starting simultaneously:
    \FOR{$i=1,2,3,\dots$}
      \STATE Record $N_\mathrm{s}$ new samples of $\dot{\mathbf{y}}_k[n]$ since $i-1$;
      \STATE WOLA analysis on $N$ most recent $\dot{\mathbf{y}}_k[n]$ samples to obtain $\mathbf{y}_k[\nu,i]$;
      \STATE Perform signal fusion via~\eqref{eq:compressed_signal_danse} to obtain $z_k[\nu,i]$;
      \STATE WOLA synthesis on $z_k[\nu,i]$ and overlap-add with previous frame to obtain $N_\mathrm{s}$ new $\dot{z}_k[n]$ samples;
      \STATE Transmit $N_\mathrm{s}$ most recent $\dot{z}_k[n]$ samples to $\mathcal{K}_k$;
      \STATE Build $\dot{\mathbf{z}}_{-k}[n]$ from samples received from $\mathcal{K}_k$;
      \STATE WOLA analysis on $N$ most recent $\dot{\mathbf{z}}_{-k}[n]$ samples to obtain $\mathbf{z}_{-k}[\nu,i]$;
      \STATE Build $\tilde{\mathbf{y}}_k[\nu,i] = [\mathbf{y}_k^T[\nu,i] \:|\: \mathbf{z}_{-k}^T[\nu,i]]^T$;
      \STATE Compute $\tilde{\mathbf{R}}_{\mathbf{y}_k\mathbf{y}_k}[\nu,i]$ and $\tilde{\mathbf{R}}_{\mathbf{n}_k\mathbf{n}_k}[\nu,i]$ via~\eqref{eq:VADexpavg};
      \STATE Compute the filter estimates $\tilde{\mathbf{w}}_k[\nu,i+1]$ via~\eqref{eq:danse_filter_update};
      \STATE Compute new frame $\hat{d}_k[\nu,i]$;
      \STATE WOLA synthesis on $\hat{d}_k[\nu,i]$ and overlap-add with the previous frame to build $\dot{\hat{d}}_k[n]$.
    \ENDFOR
  \end{algorithmic}
\end{algorithm}

In the presence of SROs, the time misalignments between the local microphone signals and the fused microphone signals from other nodes lead to incorrect covariance matrix updates which, in turn, inhibit the computation of useful filter estimates via~\eqref{eq:danse_filter_update}.
In the following section, we propose a method for SRO estimation and compensation applicable to the WOLA implementation of the DANSE algorithm.

\section{SRO estimation and compensation}\label{sec:sroestcomp}

The SRO between node $k$ and $q$ is denoted by $\varepsilon_{kq}$ such that $f_{\mathrm{s},q} = f_{\mathrm{s},k}(1 + \varepsilon_{kq})$, where $f_{\mathrm{s},k}$ and $f_{\mathrm{s},q}$ are the sampling rate of node $k$ and $q$, respectively.
In the following, it is assumed that the SROs are time-invariant and that signals recorded by the same node are synchronised.
Although a fully connected WASN is assumed, this SRO estimation and compensation method can be also adopted in other network topologies.

\subsection{Coherence-drift-based SRO estimation}\label{subsec:cohdrift}

In order to allow any node $k\in\mathcal{K}$ to blindly estimate the SROs $\{\varepsilon_{kq}\}_{q\in\mathcal{K}_k}$ using the signals it can access in the DANSE algorithm, we use a coherence-drift method based on principles introduced in~\cite{schmalenstroeer_multi-stage_2017} and~\cite{gburrek_synchronization_2022}.
At frame $i$ and at node $k$, considering one other node $q\in\mathcal{K}_k$, the available STFT-domain signals are (i) the local microphone signals $\mathbf{y}_k[\nu,i]$ and (ii) the received fused signal $z_q[\nu,i]$. The first local microphone signal $y_{k,1}[\nu,i]$ is used in the following (w.l.o.g.).

The sampling rate mismatch can simply be approximated in the STFT-domain via the linear phase drift (LPD) model~\cite{miyabe_blind_2013, wang_correlation_2016} at any frequency bin $\nu\in\{1,\dots,N\}$, i.e.:

\begin{equation}\label{eq:lpd}
  \check{z}_q[\nu,i] \approx z_q[\nu,i]\cdot\mathrm{exp}\left(\mathrm{j}\frac{2\pi}{N}\nu\varepsilon_{kq}[i] N_\mathrm{c}[i]\right),
\end{equation}

\noindent
where $\check{z}_q[\nu,i]$ is the $\varepsilon_{kq}[i]$-compensated version of $z_q[\nu,i]$ (synchronised with $y_{k,1}[\nu,i]$) and $N_\mathrm{c}[i]$ is the central sample index of frame $i$.
The product $\tau_{kq}[i] = \varepsilon_{kq}[i] N_\mathrm{c}[i]$ is the average accumulated time-drift between $z_q[\nu,i]$ and $y_{k,1}[\nu,i]$.
The LPD model relies on the assumption that the SRO-induced time drift is constant within one frame, implying that the model best approximates the effect of SROs for small $\varepsilon_{kq}[i]$.

Based on~\eqref{eq:lpd}, $\varepsilon_{kq}[i]$ can be estimated by node $k$ as follows. First, we define the instantaneous estimate of the cross-power spectral density (PSD) $\Psi_{kq}[\nu,i]$ as $\Psi_{kq}[\nu,i] = y_{k,1}[\nu,i]\cdot z_q^\ast[\nu,i]$, where $\cdot^\ast$ denotes complex conjugation.
Similarly, the instantaneous auto-PSD estimates are defined as $\Psi_{kk}[\nu,i] = |y_{k,1}[\nu,i]|^2$ and $\tilde{\Psi}_{qq}[\nu,i] = |z_q[\nu,i]|^2$.
An instantaneous estimate of the coherence between $y_{k1}[\nu,i]$ and $z_q[\nu,i]$ can then be obtained as:

\begin{equation}\label{eq:coh}
  \Gamma_{kq}[\nu,i] = \frac{\Psi_{kq}[\nu,i]}{\sqrt{\Psi_{kk}[\nu,i]\cdot \tilde{\Psi}_{qq}[\nu,i]}}.
\end{equation}

The SRO can now be estimated by defining the product $P_{\Gamma,kq}[\nu,i]$ between the instantaneous coherence estimate at frame $i$ and at frame $i-l_\mathrm{d}$ as:

\begin{equation}\label{eq:cohprod}
  P_{\Gamma,kq}[\nu,i]
  = \Gamma_{kq}[\nu,i]\cdot \Gamma_{kq}^{\ast}[\nu,i-l_\mathrm{d}].
\end{equation}

\noindent
Based on the LPD model and assuming static sources, it can be shown that an SRO estimate $\hat{\varepsilon}_{kq}[i]$ proportional to the phase of $P_{\Gamma,kq}[\nu,i]$~\cite{schmalenstroeer_multi-stage_2017, gburrek_synchronization_2022} is obtained as:

\begin{equation}\label{eq:phaseofP}
  \angle\left\{P_{\Gamma,kq}[\nu,i]\right\}
  = \frac{2\pi}{N}\nu l_\mathrm{d}N_\mathrm{s} \hat{\varepsilon}_{kq}[i],
\end{equation}

\noindent
where $\angle\{\cdot\}$ denotes the phase.
Increasing the value of $l_\mathrm{d}$ is equivalent to estimating the average SRO over a longer period of time, setting a trade-off between robust estimation and the ability to track time-varying SROs. Since fixed SROs are considered here, $l_\mathrm{d}$ may be safely set to a relatively large value, bearing in mind that SRO estimation can only begin after $l_\mathrm{d}$ frames.
Temporal averaging can be applied before computing the phase to smoothen the estimation:

\begin{equation}\label{eq:PexpAvg}
  \bar{P}_{\Gamma,kq}[\nu,i] = \alpha \bar{P}_{\Gamma,kq}[\nu,i-1] + (1 - \alpha) P_{\Gamma,kq}[\nu,i],
\end{equation}

\noindent
where $\alpha$ is a scalar, $0\ll\alpha\leq 1$, set close to 1.

Since~\eqref{eq:phaseofP} and~\eqref{eq:PexpAvg} are defined for all frequency bins $\nu$, the SRO can be estimated, for example, as the least squares (LS) solution over all relevant frequency bins~\cite{bahari2017blind}.
Since this LS solution is, however, prone to inaccuracies due to the periodicity of the phase, it has been proposed in~\cite{gburrek_synchronization_2022} to interpret $\bar{P}_{\Gamma,kq}[\nu,i]$ as a generalised cross-PSD. The integer time lag $\lambda_\mathrm{max}[i]$ that maximises the absolute value of the generalised cross-correlation $\bar{p}_{\Gamma,kq}^i[\lambda] = \mathcal{F}^{-1}\{\bar{P}_{\Gamma,kq}[\nu,i]\}$, with $\mathcal{F}^{-1}\{\cdot\}$ denoting the inverse DFT, can then be used to estimate the SRO as:

\begin{equation}\label{eq:sroest}
  \hat{\varepsilon}_{kq}[i] 
  = -\frac{\lambda_\mathrm{max}[i]}{l_\mathrm{d}N_\mathrm{s}}
  = -\frac{1}{l_\mathrm{d}N_\mathrm{s}}\cdot \underset{\lambda}{\argmax} |\bar{p}^i_{\Gamma,kq}[\lambda]|.
\end{equation}

Higher SRO estimation accuracy can be obtained by determining the non-integer value $\lambda[i]$ that maximises $|p_{\Gamma,kq}^i[\lambda]|$, via an interpolation method such as a golden section search in the interval $[\lambda_\mathrm{max}[i] - 0.5, \lambda_\mathrm{max}[i] + 0.5]$, as proposed in~\cite{gburrek_synchronization_2022}, and substituting $\lambda_\mathrm{max}[i]$ by $\lambda[i]$ in~\eqref{eq:sroest}.

\subsection{SRO compensation and full-sample drifts}

The SRO estimates obtained via the method described in Section~\ref{subsec:cohdrift} are now used to perform SRO compensation on the elements of $\mathbf{z}_{-k}[\nu,i]$ before updating $\tilde{\mathbf{w}}_k[\nu,i]$ as described in Section~\ref{sec:danse}. 
Using the LPD model, SRO compensation can be performed at any node $k$ based on $\{\hat{\varepsilon}_{kq}[i]\}_{q\in\mathcal{K}_k}$ by applying the appropriate phase shift to each element of $\mathbf{z}_{-k}[\nu,i]$ as:

\begin{equation}\label{eq:SROcomp_phaseShifts}
  \check{z}_q[\nu,i] = z_q[\nu,i] \cdot \mathrm{exp}\left(\mathrm{j}\frac{2\pi}{N}\nu \hat{\tau}_{kq}[i]\right) \:\forall\:q\in\mathcal{K}_k,
\end{equation}

\noindent
where $\hat{\tau}_{kq}[i] = N_\mathrm{s} \sum_{\iota=0}^i \hat{\varepsilon}_{kq}[\iota]$ is the estimated average accumulated time-drift between $z_q[\nu,i]$ and $y_{k1}[\nu,i]$ (cfr.~\eqref{eq:lpd}).

\begin{figure}[h]
  \centering
  \includegraphics[width=\columnwidth,trim={0 3em 0 0},clip=false]{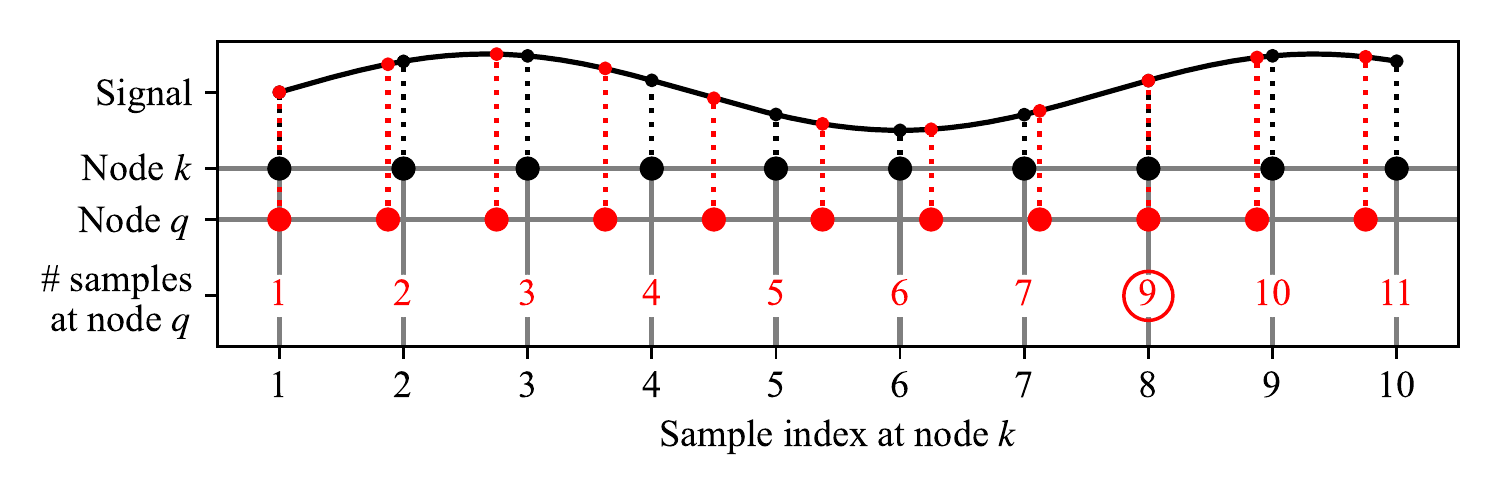}
  \caption{Schematic representation of a full-sample drift (indicated by the circle) generated by an SRO $\varepsilon_{kq} > 0$ between node $k$ and $q\in\mathcal{K}_k$.}
  \label{fig:fullsampledrift}
\end{figure}

An important aspect comes into play $t^\mathrm{FSD}_{kq} = 1/(f_{\mathrm{s},k}|\varepsilon_{kq}|)$ seconds after the simultaneous initialisation of the WASN, namely when the accumulated SRO-induced time drift $\tau_{kq}[i]$ between node $k$ and node $q$ becomes greater than one sample.
Such event is referred to in the following as a full-sample drift (FSD).
At that time, if $\varepsilon_{kq} > 0$, the growing time drift between node $k$ and node $q$ leads to a situation where node $q$ has recorded one more sample than node $k$, as depicted in Figure~\ref{fig:fullsampledrift}. Conversely, if $\varepsilon_{kq} < 0$, node $q$ has recorded one less sample than node $k$.

When correctly detected, an FSD can be compensated for by applying a corrective phase shift $\phi_{kq}^{\mathrm{FSD}}[\nu,i]$ to $z_q[\nu,i]$ as:

\begin{equation}\label{eq:flag_phaseShift}
  \phi_{kq}^{\mathrm{FSD}}[\nu,i] =
  \begin{cases}
    \mathrm{exp}\left(-\mathrm{j}\frac{2\pi}{N}\nu\right)& \text{if one more sample at $q$,}\\
    \mathrm{exp}\left(\mathrm{j}\frac{2\pi}{N}\nu\right)& \text{if one less sample at $q$,}\\
    1              & \text{otherwise.}
  \end{cases}
\end{equation}

The SRO estimation itself can be biased by the presence of one or more FSDs between frame $i - l_\mathrm{d}$ and frame $i$. These can be accounted for by multiplying $P_{\Gamma,kq}[\nu,i]$ by the accumulated FSD phase shift:

\begin{equation}\label{eq:accFSD_shift}
  \phi_{kq}^{\mathrm{ac}}[\nu,i] = \prod_{\iota=i-l_\mathrm{d}}^i \phi^{\mathrm{FSD}}_{kq}[\nu,\iota].
\end{equation}
\noindent
The rest of the SRO estimation process remains unchanged, following~\eqref{eq:PexpAvg} and~\eqref{eq:sroest}.

\begin{figure}[htb]
  \centering
  \centerline{
    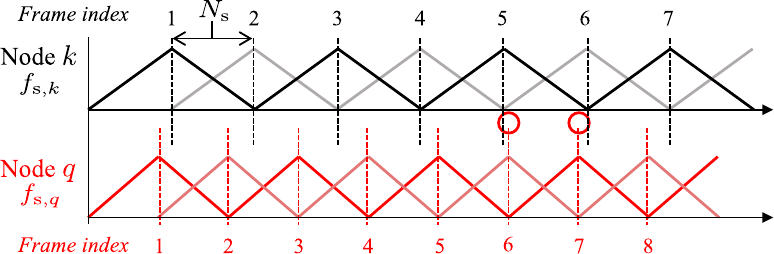
  }
  \caption{Schematic representation of a full-frame drift (highlighted by circles), with $\varepsilon_{kq} > 0$ and 50\% WOLA window shift.}
  \label{fig:ffds}
\end{figure}

However, the accumulated effect of FSDs becomes particularly problematic when considering the WOLA implementation of DANSE~\cite{bertrand_robust_2009}, where a fused time-domain signal $\dot{z}_q[n]$ is transmitted in frames of $N_\mathrm{s}$ samples from node $q$ to node $k$ (cfr.~Algorithm~\ref{alg:woladanse}). For clarity of exposition, we assume an even DFT size $N$ and a 50\% WOLA window shift such that $N_\mathrm{s} = N/2$. A problematic phenomenon referred to as full-frame drift (FFD) occurs when $N_\mathrm{s}$ uncompensated FSDs accumulate. If $\varepsilon_{kq} > 0$ (resp. $\varepsilon_{kq}<0$) and after $t^\mathrm{FFD}_{kq} = N_\mathrm{s} t^\mathrm{FSD}_{kq}$ seconds, node $q$ has recorded $N_\mathrm{s}$ more (resp. less) samples than node $k$ since the synchronous initialisation of both nodes. At that time, node $q$ has thus transmitted \textit{two} (resp. \textit{no}) new $\dot{z}_q[n]$ frames since the last update of node $k$ (see circles on Figure~\ref{fig:ffds}). Consequently, to perform its next update, node $k$ \textit{skips} (resp. \textit{duplicates}) one $z_q[\nu,i]$ frame.

An FFD cannot be compensated for via a phase shift based on~\eqref{eq:flag_phaseShift} if $N_\mathrm{s}$ is close to $N$.
For instance, with 50\% WOLA window shift, the corrective phase shift of~\eqref{eq:flag_phaseShift} needed to compensate for $N_\mathrm{s}$ FSDs at once simplifies to $\mathrm{exp}(\pm\mathrm{j}\frac{2\pi}{N}\nu N_\mathrm{s}) = \pm 1$ $\forall\:\nu$. Even if FFDs compensation were possible, before an FFD occurs node $k$ receives a single $N_\mathrm{s}$ samples-long frame of $\dot{z}_q[n]$ between two consecutive filter updates, as in Figure~\ref{fig:ffds}. Node $k$ is, therefore, unable to detect FSDs by comparing the number of local $\dot{\mathbf{y}}_k[n]$ samples with the number of received $\dot{z}_q[n]$ samples since its previous update (as both are equal to $N_\mathrm{s}$). The uncompensated growing drift between elements of $\tilde{\mathbf{y}}_k[\nu,i]$ then leads to increasingly erroneous updates of the covariance matrices.

If disregarded, FFDs can significantly perturb the convergence of DANSE as well as the SRO estimation process. 
The detection of FSDs within the WOLA implementation of DANSE is discussed in the following section.

\subsection{Full-sample drift detection}\label{subsec:fullsample_drifts}

In order to enable detection and compensation of FSDs within the WOLA implementation of DANSE, we introduce a modification of the DANSE fusion and broadcasting mechanism to allow per-sample transmission of fused signals between nodes, while retaining WOLA frame-by-frame processing for the computationally costly steps of GEVD-based filter update and desired signal estimate computation. 
In principle, using this per-sample transmission, node $k$ can easily detect FSDs for the $i$-th filter update by comparing the number of local $\dot{\mathbf{y}}_k[n]$ samples with the number of received $\dot{z}_q[n]$ samples from node $q$ since its previous update, then compensate for them via the corrective phase shifts of~\eqref{eq:flag_phaseShift}.

We propose to approximate the WOLA filtering process (analysis, STFT-domain filtering, and synthesis) by its so-called distortion function $T(\zeta)$~\cite{vaidyanathan2006multirate}, where $\zeta$ is the $\mathcal{Z}$-transform variable. This function relates the output of the WOLA filterbank to its input when no decimation and expansion is performed, i.e., using maximal window overlap.
At frame $i$, the distortion function $T^i_{q,m}(\zeta)$ corresponding to the $m$-th microphone of node $q$ can be obtained as:

\begin{equation}\label{eq:distfcn}
  T_{q,m}^i(\zeta)
  = 
  \frac{1}{N_\mathrm{s}}
  \begin{bmatrix}
      \zeta^{1-N} \,\dots\, 1
  \end{bmatrix}
  \mathbf{D}_{q,m}^i
  \begin{bmatrix}
      1 \,\dots\, \zeta^{1-N}
  \end{bmatrix}^T,
\end{equation}

\noindent
with $\mathbf{D}_{q,m}^i
  =
  \mathbf{H}_\mathrm{s}
  \cdot 
  \mathbf{F}^{-1}
  \cdot 
  \mathrm{diag}\{\mathbf{w}_{qq,m}[i]\}
  \cdot 
  \mathbf{F}
  \cdot 
  \mathbf{H}_\mathrm{a}$,
where $\mathbf{F}^{-1}$ and $\mathbf{F}$ are the inverse DFT and DFT matrix, respectively, $\mathbf{w}_{qq,m}[i] =\begin{bmatrix}
  w_{qq,m}[1,i],\dots,w_{qq,m}[N,i]
\end{bmatrix}^T$ denotes the local filter coefficients at frame $i$
for the $m$-th microphone of node $q$ with all frequency bins stacked into one vector, $\mathbf{H}_\mathrm{s} = \mathrm{diag}\{\mathrm{flip}\{\mathbf{h}_\mathrm{s}\}\}$, and $\mathbf{H}_\mathrm{a} = \mathrm{diag}\{\mathbf{h}_\mathrm{a}\}$, respectively, where $\mathbf{h}_\mathrm{s}$ and $\mathbf{h}_\mathrm{a}$ denote the WOLA synthesis and analysis time-domain windows, respectively, and $\mathrm{flip}\{\cdot\}$ reverses the order of the elements of a vector.

The time-domain equivalent of the distortion function $T_{q,m}^i(\zeta)$ in~\eqref{eq:distfcn} is a $(2N-1)$-tap impulse response denoted by $\mathbf{t}_{q,m}^i$. From~\eqref{eq:distfcn}, it can be seen that each element of $\mathbf{t}_{q,m}^i$ is obtained by summing over the corresponding diagonal of the matrix $\mathbf{D}_{q,m}^i$.
The complete WOLA analysis and synthesis process can then be approximated by a convolution with $\mathbf{t}_{q,m}^i$.
This means that the $n$-th sample of the time-domain fused signal $\dot{z}_q[n]$ can be obtained as:

\begin{equation}\label{eq:fusionTD}
  \dot{z}_q[n] = \sum_{m=1}^{M_q} \left(\dot{\mathbf{y}}_{q,m}^{(n)} \ast \mathbf{t}_{q,m}^i\right)[n+2N-1]
\end{equation}

\noindent
where the time-domain vector $\dot{\mathbf{y}}_{q,m}^{(n)}$ contains the most recent $N$ samples recorded by the $m$-th microphone of node $q$ and $(\mathbf{a}\ast \mathbf{b})[c]$ denotes the $c$-th sample of the convolution between time-domain signals $\mathbf{a}$ and $\mathbf{b}$.
Note that the distortion function does not need be computed at every frame $i$, especially once the filters have converged after several DANSE iterations. The iteration indices at which the distortion function is updated with the most recent filter $\mathbf{w}_{qq}[\nu,i]$ are grouped in the set $\mathcal{I}_\mathrm{T}$.

Although the proposed $T(\zeta)$-approximation introduces the same $N-1$ samples input-output delay as the standard WOLA implementation of DANSE~\cite{bertrand_robust_2009}, it has the advantage to circumvent the $N_\mathrm{s}$ samples delay introduced by frame-by-frame processing~\cite{crochiere_weighted_1980} since no downsampling is performed. Additionally, the use of per-sample broadcasting reduces the amount of transmitted data as each compressed signal sample is transmitted only once. This differs from the usual WOLA scheme where the overlap-add operation necessitates the transmission of $N_\mathrm{s}$ additional data points per $N$-samples block of compressed signal (as in~Algorithm~\ref{alg:woladanse}).

\begin{figure}[htb]
  \centering
  \centerline{
    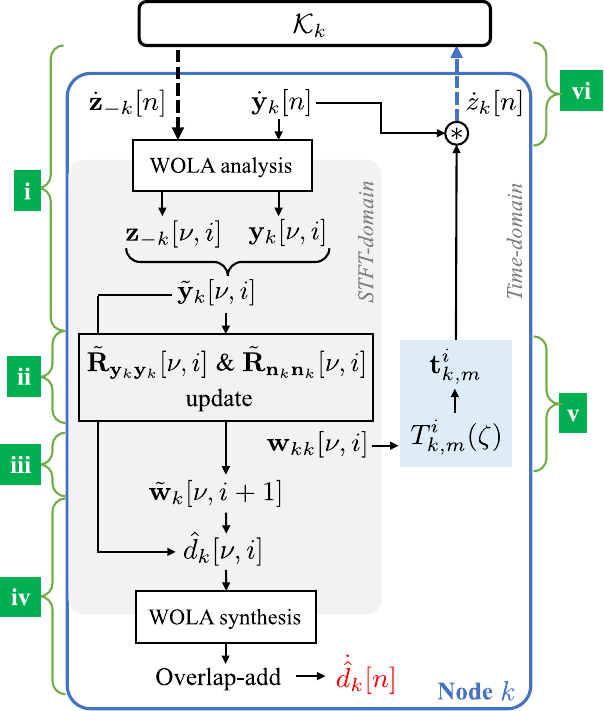
  }
  \caption{Proposed WOLA-based DANSE processing at node $k\in\mathcal{K}$ with per-sample fused signal broadcasting. 
  [i] WOLA analysis applied to local microphone signals $\dot{\mathbf{y}}_k[n]$ and fused signals from other nodes $\dot{\mathbf{z}}_{-k}[n]$. [ii] Covariance matrix update and [iii] computation of filter $\tilde{\mathbf{w}}_k[\nu,i+1]$. [iv] Computation of new desired signal estimate frame $\hat{d}_k[\nu,i+1]$, followed by WOLA synthesis and overlap-add. [v] Computation of distortion functions $\{T_{k,m}^i(\zeta)\}_{m=1}^{M_k}$ from filter $\mathbf{w}_{kk}[\nu,i]$. [vi] Computation of new $\dot{z}_k[n]$ samples and per-sample broadcasting to other nodes.}
  \label{fig:woladanseblock}
\end{figure}

Using the $T(\zeta)$-approximation, any node is able to broadcast its fused signal on a per-sample basis. This modification of the DANSE algorithm, although coming at the expense of some additional computational complexity with respect to a purely WOLA-based implementation, enables the deployment of DANSE in asynchronous WASNs where FSDs can be detected as soon as they occur.
An overview of the DANSE algorithm with per-sample fused signal broadcasting using the $T(\zeta)$-approximation is provided in Figure~\ref{fig:woladanseblock}.

\subsection{Complete system}\label{subsec:complete_system}

\begin{figure}[htb]
  \centering
  \centerline{
    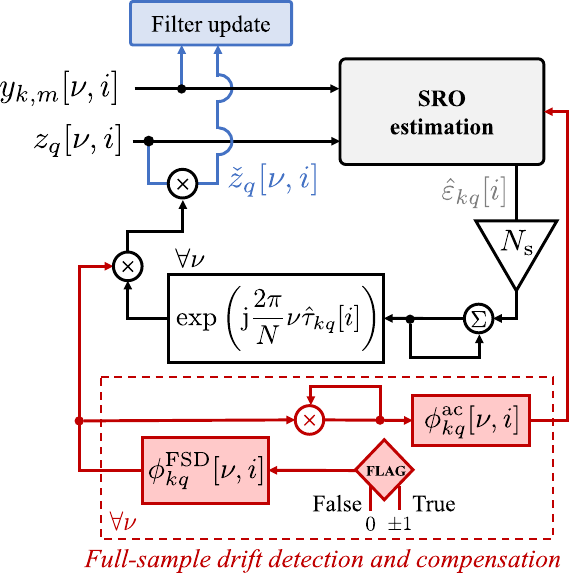
  }
  \caption{SRO estimation and compensation block-scheme at node $k\in\mathcal{K}$, including full-sample drift detection (``flag'') and compensation.}
  \label{fig:sroestcomp}
\end{figure}

As SRO estimation is necessary for SRO compensation, both should be performed in parallel.
An open-loop strategy is proposed, as depicted in Figure~\ref{fig:sroestcomp}, which consists of three parts: SRO estimation, FSD detection, and SRO compensation. 
First, the SRO-uncompensated fused signal $z_q[\nu,i]$ is used to estimate $\hat{\varepsilon}_{kq}[i]$. 
Every time an FSD is detected, a flag is raised and the FSD phase shift of~\eqref{eq:flag_phaseShift} is included when performing SRO estimation and compensation, leading to the signal $\check{z}_q[\nu,i]$, which is used to update the DANSE filter.

Algorithm~\ref{alg:woladanse_sros} provides a complete description of WOLA-based DANSE with SRO estimation and compensation, including the $T(\zeta)$-approximation for FSD detection. 
There, the STFT-domain SRO-compensated fused signals vector is denoted by $\check{\mathbf{z}}_{-k}[\nu,i]$ and the SRO-compensated version of $\tilde{\mathbf{y}}_k[\nu,i]$ becomes $\check{\mathbf{y}}_k[\nu,i] = [\mathbf{y}_k^T[\nu,i] \:|\: \check{\mathbf{z}}_{-k}^T[\nu,i]]^T$. The estimates of $E\{\check{\mathbf{y}}_k[\nu,i]\check{\mathbf{y}}_k^H[\nu,i]\}$ and $E\{\check{\mathbf{n}}_k[\nu,i]\check{\mathbf{n}}_k^H[\nu,i]\}$ are denoted by $\check{\mathbf{R}}_{\mathbf{y}_k\mathbf{y}_k}[\nu,i]$ and $\check{\mathbf{R}}_{\mathbf{n}_k\mathbf{n}_k}[\nu,i]$, respectively. The filter estimate after SRO compensation is finally obtained similarly to~\eqref{eq:danse_filter_update}, i.e., performing a GEVD on the matrix pencil $\{\check{\mathbf{R}}_{\mathbf{y}_k\mathbf{y}_k}[\nu,i], \check{\mathbf{R}}_{\mathbf{n}_k\mathbf{n}_k}[\nu,i]\}$, and is denoted by $\check{\mathbf{w}}_k[\nu,i+1]$. 

\begin{algorithm}
  \caption{WOLA-based DANSE with SRO compensation in a fully connected heterogeneous WASN.}\label{alg:woladanse_sros}
  \begin{algorithmic}[1]
    \STATE Initialise $\tilde{\mathbf{w}}_k[\nu,0]$ $\forall\:(k,\nu)\in\mathcal{K}\times\{1,\dots,N\}$;
    \STATE Each node $k\in\mathcal{K}$ performs, starting simultaneously:
    \FOR{every new locally recorded sample $\dot{\mathbf{y}}[n]$}
      \STATE Compute $\dot{z}_k[n]$ via~\eqref{eq:fusionTD} and transmit to nodes in $\mathcal{K}_k$.
    \ENDFOR
    \FOR{$i=1,2,3,\dots$}
      \IF{$i\in\mathcal{I}_\mathrm{T}$}
        \STATE Update $\{T_{k,m}^i(\zeta)\}_{m=1}^{M_k}$ via~\eqref{eq:distfcn} using $\mathbf{w}_{kk}[\nu,i]$;
      \ELSE
        \STATE $\{T_{k,m}^i(\zeta)\}_{m=1}^{M_k} = \{T_{k,m}^{i-1}(\zeta)\}_{m=1}^{M_k}$;
      \ENDIF
      \STATE Shift WOLA window ($N_\mathrm{s}$ new samples since $i-1$);
      \STATE WOLA analysis on local signals to obtain $\mathbf{y}_k[\nu,i]$;
      \STATE WOLA analysis on fused signals to obtain $\mathbf{z}_{-k}[\nu,i]$;
      \FOR{$q\in\mathcal{K}_k$}
        \STATE Detect FSDs based on number of new $\dot{z}_q[n]$ samples and compute $\phi_{kq}^{\mathrm{ac}}[\nu,i]$ $\forall\:\nu$ via~\eqref{eq:flag_phaseShift} and \eqref{eq:accFSD_shift};
        \STATE Compute $\hat{\varepsilon}_{kq}[i]$ via~\eqref{eq:sroest};
        \STATE Compute $\check{z}_q[\nu,i]$ (Figure~\ref{fig:sroestcomp}) and build $\check{\mathbf{z}}_{-k}[\nu,i]$;
      \ENDFOR
      \STATE Build $\check{\mathbf{y}}_k[\nu,i] = [\mathbf{y}_k^T[\nu,i] \:|\: \check{\mathbf{z}}_{-k}^T[\nu,i]]^T$;
      \STATE Compute $\check{\mathbf{R}}_{\mathbf{y}_k\mathbf{y}_k}[\nu,i]$ and $\check{\mathbf{R}}_{\mathbf{n}_k\mathbf{n}_k}[\nu,i]$;
      \STATE Compute $\check{\mathbf{w}}_k[\nu,i+1]$ via~\eqref{eq:danse_filter_update}, then $\hat{d}_k[\nu,i]$.
      \STATE WOLA synthesis on $\hat{d}_k[\nu,i]$ and overlap-add with the previous frame to build $\dot{\hat{d}}_k[n]$.
    \ENDFOR
  \end{algorithmic}
\end{algorithm}

\section{Numerical experiments}\label{sec:results}

\begin{figure}[htb]
  \centering
  \includegraphics[width=.9\columnwidth,trim={0 3em 0 0},clip=false]{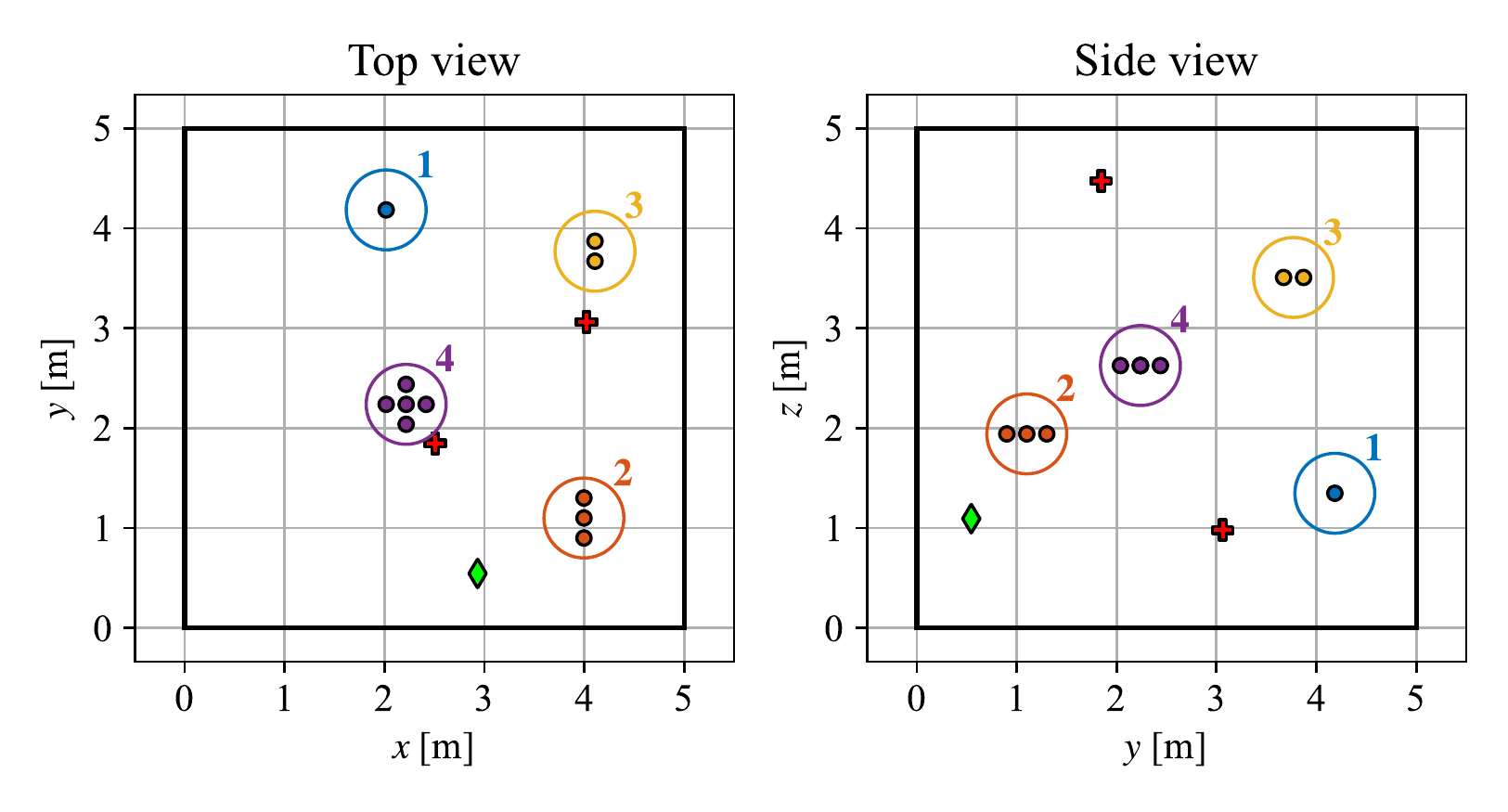}
    \caption{Layout of acoustic scenario used in the simulations. Microphones ($\circ$) are grouped in nodes ($\bigcirc$) numbered 1 to 4. Two noise sources (+) and one desired source ($\diamond$) are present.}
  \label{fig:acoustic_scenario}
\end{figure}

The performance of Algorithm~\ref{alg:woladanse_sros} is demonstrated and compared to Algorithm~\ref{alg:woladanse} via numerical experiments. The acoustic environment is depicted in Figure~\ref{fig:acoustic_scenario}. 
A WASN of $K=4$ nodes is considered, with $\{M_k\}_{k=1}^4 = \{1,3,2,5\}$ microphones with a 20 cm inter-microphone spacing.
A 5$\times$5$\times$5 m\textsuperscript{3} room with a uniform absorption coefficient of 0.9 is considered, resulting in a $T_{60} = 0.15$ s reverberation time. One localised speech source and two localised uncorrelated stationary white noise sources are present (note that the validity of Algorithm~\ref{alg:woladanse_sros} can also be demonstrated in the presence of a non-stationary noise source such as babble noise).
The speech signal consists of 3 s long LibriSpeech~\cite{panayotov2015librispeech} snippets, each separated by 2 s of silence and starting with 0.25 s of silence.
The power of each source is set to obtain a -3 dB signal-to-noise ratio (SNR) at the reference microphone of node 1.
All signals last 15 s and are simulated by convolving the source signals with 4096 samples room impulse responses obtained using the randomised image method~\cite{de_sena_modeling_2015}. The nominal sampling rate is set to 16 kHz.

The filters are initialised as selecting the local reference microphone signal, i.e., $\tilde{\mathbf{w}}_k[\nu,0] = [1 \: \mathbf{0}]^T\:\forall\:k\in\mathcal{K}$. The covariances matrices are updated using $\beta=0.978$ (cfr.~\eqref{eq:VADexpavg}). All WOLA processing is performed using $N=1024$-samples square-root Hann windows with 50\% overlap (note that the conclusions presented here in terms of speech enhancement are also valid for other frame lengths, e.g., $N=512$ or $2048$ samples).
FSDs are detected using the WOLA approximation described in Section~\ref{subsec:fullsample_drifts}, where the distortion function $T_{k,m}^i(\zeta)$ in~\eqref{eq:distfcn} is updated based on the filter $\mathbf{w}_{kk}[\nu,i]$ every 30 DANSE iterations.
The covariance matrices are estimated via~\eqref{eq:VADexpavg} assuming an ideal VAD, which avoids the influence of VAD errors on the results. In practice, the VAD obviously needs to be estimated from the microphone signals~\cite{bertrand_energy-based_2010,zhao_model-based_2020}.

The clock of node 1 is set as the reference, with $f_{\mathrm{s},1}=16$ kHz.
The SRO for all other nodes $k\in\{2,3,4\}$ is defined with respect to this reference. Three degrees of network asynchronicity are considered based on the measured SROs values reported in~\cite{he_analysis_2015}. First, small SROs are considered by setting $\{\varepsilon_{1k}\}_{k=2}^4 = \{20,-20,40\}$ PPM. Second, more asynchronicity is applied by setting $\{\varepsilon_{1k}\}_{k=2}^4 = \{50,-50,100\}$ PPM. Finally, a strongly asynchronous network is simulated by setting $\{\varepsilon_{1k}\}_{k=2}^4 = \{200,-200,400\}$ PPM. Fixed SROs are simulated at any node by resampling the signals appropriately. The SRO estimation method uses $l_\mathrm{d} = 10$ in~\eqref{eq:cohprod} and $\alpha=0.95$ in~\eqref{eq:PexpAvg}, resulting in a $\pm$ 3 PPM accuracy.

\begin{figure}[h]
  \centering
  \includegraphics[width=\columnwidth,trim={0 3em 0 0},clip=false]{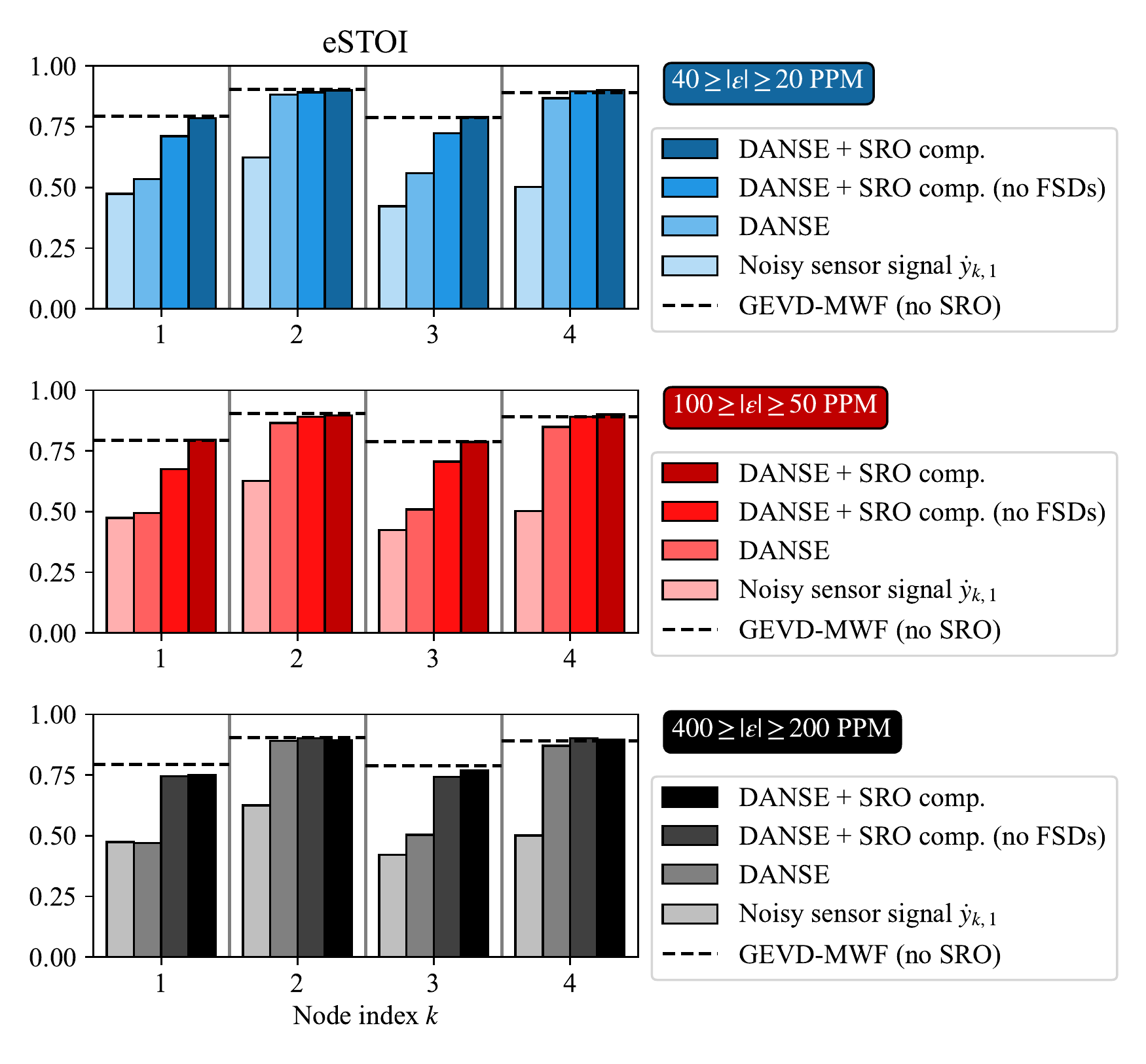}
  \caption{eSTOI obtained at each node from Figure~\ref{fig:acoustic_scenario} with small SROs (top), moderate SROs (middle), or large SROs (bottom). Local noisy reference microphone signals without any processing (lightest), DANSE estimates in the presence of SROs (light), DANSE estimates with SRO estimation and compensation, without FSD compensation (dark) and with (darkest), and MWF SRO-free centralised estimates (dashed).}
  \label{fig:results}
\end{figure}

The performance at each node is quantified using the extended short-term objective intelligibility (eSTOI)~\cite{jensen2016algorithm} with the clean speech component of the first local microphone as reference. This metric is particularly relevant as opposed to, e.g., SNR, as intelligible speech is of central interest in most speech enhancement applications.
The eSTOI is computed on the signal segment starting from WOLA frame $i=15$ to reduce the impact of initial filter updates.
For each degree of asynchronicity, Figure~\ref{fig:results} shows the eSTOI at each node for the local reference microphone signal (without any noise reduction), the desired signal estimate from Algorithm~\ref{alg:woladanse} without SRO compensation, and the desired signal estimate from Algorithm~\ref{alg:woladanse_sros} with the proposed SRO compensation with or without compensating for FSDs.
The eSTOI obtained using the synchronised and centralised GEVD-MWF (cfr.~\eqref{eq:gevd-mwf}) is provided for comparison.

The results show that the presence of SROs in the WASN significantly deteriorates the performance of WOLA-based GEVD-DANSE. The single-microphone node ($M_1=1$) is particularly sensitive to the presence of SROs as it heavily relies on the information provided by other nodes to compute its desired signal estimate.
This occurs regardless of the considered SRO, which shows the negative impact of even relatively small SROs. Conversely, nodes including many microphones (e.g., $M_4=5$), show almost no sensitivity to SROs, suggesting that these nodes are able to rely solely on their locally recorded signals to perform noise reduction with a comparable performance as in the centralised case.
For all considered SRO magnitudes, each node using the proposed method with FSD compensation is able to restore the centralised performance that GEVD-DANSE would showcase in an SRO-free WASN.

\section{Conclusion}\label{sec:ccl}

In this contribution, the WOLA-based implementation of the GEVD-DANSE algorithm has been rendered robust to the presence of SROs by combining a coherence-based SRO estimation technique with an approximation of the WOLA process to allow FSDs detection and compensation via per-sample broadcasting of fused signals. The performance of the proposed method has been assessed through numerical experiments in the context of speech enhancement. in terms of intelligibility of the desired signal estimate at each node.
The results show that even relatively small SROs (if not estimated and compensated for) can have a detrimental impact on the ability of DANSE to recover the desired signal at nodes that significantly rely on collaboration with other nodes. However, it is shown that, in an asynchronous WASN, the proposed SRO estimation and compensation method practically restores the performance that the GEVD-DANSE algorithm would showcase in a fully synchronised network.

\bibliographystyle{IEEEbib_mod}
\bibliography{IEEEabrv,refs}

\end{document}